\documentclass[pra,twocolumn,showpacs,superscriptaddress,floatfix]{revtex4}
\font\tenmib=cmmib10 
\font\sevenmib=cmmib7
\font\fivemib=cmmib5 
\font\msytw=msbm9 scaled\magstep1%
\mathchardef\Ba   = "050B  
\mathchardef\Bb   = "050C  
\mathchardef\Bg   = "050D  
\mathchardef\Bd   = "050E  
\mathchardef\Be   = "0522  
\mathchardef\Bee  = "050F  
\mathchardef\Bz   = "0510  
\mathchardef\Bh   = "0511  
\mathchardef\Bthh = "0512  
\mathchardef\Bth  = "0523  
\mathchardef\Bi   = "0513  
\mathchardef\Bk   = "0514  
\mathchardef\Bl   = "0515  
\mathchardef\Bm   = "0516  
\mathchardef\Bn   = "0517  
\mathchardef\Bx   = "0518  
\mathchardef\Bom  = "0530  
\mathchardef\Bp   = "0519  
\mathchardef\Br   = "0525  
\mathchardef\Bro  = "051A  
\mathchardef\Bs   = "051B  
\mathchardef\Bsi  = "0526  
\mathchardef\Bt   = "051C  
\mathchardef\Bu   = "051D  
\mathchardef\Bf   = "0527  
\mathchardef\Bff  = "051E  
\mathchardef\Bch  = "051F  
\mathchardef\Bps  = "0520  
\mathchardef\Bo   = "0521  
\mathchardef\Bome = "0524  
\mathchardef\BG   = "0500  
\mathchardef\BD   = "0501  
\mathchardef\BTh  = "0502  
\mathchardef\BL   = "0503  
\mathchardef\BX   = "0504  
\mathchardef\BP   = "0505  
\mathchardef\BS   = "0506  
\mathchardef\BU   = "0507  
\mathchardef\BF   = "0508  
\mathchardef\BPs  = "0509  
\mathchardef\BO   = "050A  
\mathchardef\BDpr = "0540  
\mathchardef\Bstl = "053F  

\let\a=\alpha \let\b=\beta  \let\g=\gamma  \let\d=\delta \let\e=\varepsilon
\let\z=\zeta     \let\th=\theta  
\let\m=\mu                 
\let\s=\sigma \let\t=\tau   \let\f=\varphi 
     \let\o=\omega
    
    \let\Si=\Sigma     
\let\O=\Omega 

\def\\{\hfill\break} 
\def\*{\vglue0.3truecm}
\let\==\equiv
\def\tende#1{\,\vtop{\ialign{##\crcr\rightarrowfill\crcr
 \noalign{\kern-1pt\nointerlineskip} \hskip3.pt${\scriptstyle
 #1}$\hskip3.pt\crcr}}\,}
\def\otto{\,{\kern-1.truept\leftarrow\kern-5.truept\to\kern-1.truept}\,}
\def\media#1{{\langle#1\rangle}}
\def\defi{\,{\buildrel def\over=}\,}
\def\V#1{{\bf#1}}
\def\RRR{\hbox{\msytw R}}

\newcommand\revtex{{R\kern-0.24mm\lower0.5mm\hbox{E}\kern-0.6mm V\kern-0.5mm%
\lower0.5mm\hbox{T}\kern-0.4mm E\kern-.2mm \lower0.5mm\hbox{X}}}

\begin{document}

\title{Fluctuations relation and external thermostats:\\ 
an application to granular materials}
\*

\author{F.~Bonetto}
\affiliation{School of Mathematics, 
Georgia Institute of Technology, Atlanta, Georgia 30332}
\author{G.~Gallavotti}
\affiliation{Dipartimento di Fisica and INFN, Universit\`a di Roma 
{\em La Sapienza},
P.~A.~Moro 2, 00185, Roma, Italy}
\author{A. Giuliani}
\affiliation{
Department of Physics, Jadwin Hall, Princeton University, P.O. Box 708,
Princeton, New Jersey 08542
}
\author{F. Zamponi}
\affiliation{ Laboratoire de Physique Th\'eorique de 
l'\'Ecole Normale Sup\'erieure, 24 Rue Lhomond, 75231 Paris Cedex 05, France}

\date{\today}
\begin{abstract} 
In this note we discuss a paradigmatic example of 
interacting particles subject to non conservative external forces and 
to the action of thermostats consisting of external (finite) reservoirs of 
particles. We then consider a model of granular materials of interest
for experimental tests that had recently attracted lot of attentions.
This model can be reduced to the previously discussed example 
under a number of assumptions, in particular that inelasticity due to internal 
collisions can be neglected for the purpose of measuring the large deviation
functional for entropy production rate. We show that if the restitution 
coefficient in the granular material model is close to one, then the 
required assuptions are verified on a specific time scale and we predict a 
fluctuation relation for the entropy production rate measured on the same time 
scale.
\end{abstract}

\maketitle

\section{Nonequilibrium}

In studying stationary states in nonequilibrium statistical mechanics,
\cite{Ga98,Ga04}, it is common to consider $d$--dimensional
systems of particles in a
(finite) container $\Si_0$ forced by non conservative forces whose
work is controlled by thermostats consisting of particles moving
outside $\Si_0$ and interacting with the particles of $\Si_0$ through
interactions across the walls of $\Si_0$, \cite{Ru99}.
If $\V X^0=(\V x_1^0,\ldots$, $\V x_{N_0}^0)$ are the particles positions in an
inertial system of coordinates, the equations of motion are
determined by their mass $m$, by the {\it potential energy} of
interaction $V(\V X^0)$, by external
nonconservative forces $\V F_i(\V X^0,\BF)$ and by {\it thermostat}
forces $-\Bth_i$ as
\begin{eqnarray}m {\ddot{\V x}^0}_i=-\partial_{\V x_i^0} V(\V X^0)+ 
\V F_i(\V X^0; \BF)
-\Bth_i\;, \label{1.1}\end{eqnarray}
where $i=1,\ldots,N_0$ and 
$\BF=(\f_1,\ldots,\f_q)$ are strength parameters on which the
external forces depend ({\it e. g.} 
the components of an external electric field). 
Forces and potentials will be supposed
smooth, {\it i.e.} analytic, in their variables aside from {\it possible}
impulsive elastic forces describing shocks; the forces $\V F_i$ will be 
supposed to vanish for $\BF=\V 0$. 
The impulsive forces are allowed here to model
possible shocks with the walls of the container $\Si_0$ or between hard
core particles.

Examples of deterministic reservoirs, \cite{EM90}, are forces
obtained by imposing a nonholonomic constraint via some {\it ad hoc}
principle, like the Gauss' principle, \cite{Ga00} (appendix 9.A4), \cite{Ga05}.
A different example will be extensively discussed below.

In general the forces $\Bth_i$ can be considered a set of deterministic
``thermostat forces'' if a further property holds: namely that the
system evolves according to Eq. (\ref{1.1}) towards a {\it stationary
state}. This means that for all $(\dot{\V X}^0,\V X^0)$, {\it except
possibly for a set of zero phase space volume}, {\it any} smooth
function $f(\dot{\V X}^0,\V X^0)$ evolves in time so that, denoting
$S_t(\dot{\V X}^0,\V X^0)$ the configuration into which $(\dot{\V X}^0,\V
X^0)$ evolves in time $t$ according to Eq. (\ref{1.1}), then the limit
\begin{eqnarray}\lim_{T\to\infty} \frac1T \int_0^T f(S_t(\dot{\V X}^0,\V X^0))
\,dt =\int f(z)\m(dz)\label{1.2}\end{eqnarray}
exists and is independent of $(\dot{\V X}^0,\V X^0)$.  The 
phase space probability
distribution $\m(dz)$ (here $z=(\dot{\V Y},\V Y)$ denotes the coordinates of position and velocity 
of a generic point in phase space) 
is then called the {\it SRB distribution} for the
system. It has to be stressed that the condition that thermostat
forces be ``effective'' enough to impede an indefinite build up of the
energy of the system is a strong condition which we will assume in the 
models discussed in this note.
It imposes on the interaction potentials and on the thermostats conditions 
that are not well understood, although they seem empirically verified with the
simplest choices of molecular potentials, \cite{Ga00}.

The maps $S_t$ will have the group property $S_t\cdot
S_{t'}=S_{t+t'}$ and the SRB distribution $\m$ will be invariant under
time evolution. The SRB distribution is said to describe a stationary
state of the system; it depends on the parameters on which the forces
acting on the system depend, {\it e.g.} $|\Si|$ (volume), $\BF$ (strength of
the forcings) and on the model of thermostat forces. 
The collection of SRB distributions obtained by
letting the parameters vary defines a {\it nonequilibrium ensemble}.

\section{A model of external thermostats}

Let us now discuss in more detail an important  
class of thermostats in the context of model Eq. (\ref{1.1}). Imagine that the
$N_0$ particles in the container $\Si_0$ interacting via the potential
$V(\V X^0)=\sum_{i<j} \f (\V x_i^0-\V x_j^0)+\sum_j V_0(\V x_j^0)$ (where 
$V_0$ models external conservative forces like obstacles, walls, gravity, 
$\ldots$) and subject to the external forces $\V F_i$ 
are also interacting with $M$ other systems $\Si_a$, of $N_a$
particles of mass $m_a$, in containers $\Si_a$ contiguous to $\Si_0$. It will 
be assumed that $\Si_a\cap\Si_{a'}=\emptyset$ for $a\neq a'$, $a,a'=0,
\ldots,M$.

The coordinates of the particles in the $a$-th system $\Si_a$ will be
denoted $\V x^a_j, \, j=1,\ldots,N_a$, and they will interact with
each other via a potential $V_a(\V X^a)=\sum_{i,j}^{N_a}
\f_a(\V x^a_i-\V x^a_j)$. Furthermore there will be an interaction
between the particles of each thermostat and those of the system via
potentials $W_a(\V X^0,\V X^a)=\sum_{i=1}^{N_0}\sum_{j=1}^{N_a} w_a(\V x_i^0-\V
x^a_j)$, $a=1,\ldots,M$.

Again, potentials will be assumed to be either hard core or non singular
and $V_0$ to be at least such that it forbids existence of obvious constants 
of motion.

The temperature of each $\Si_a$ will be defined by the total kinetic
energy of its particles {\it i.e.}, $k_B$ being Boltzmann's constant,
by setting $K_a=\sum_{j=1}^{N_a} \frac12 m_a (\dot{\V x}^a_j)^2\defi
\frac12 (dN_a-1) k_B T_a$, where $d$ is the spatial dimension;
the particles of the $a$-th thermostat will be
kept at constant temperature by further forces $\Bth^a_j$.  The latter
are defined by imposing constancy of $K_a$ via Gauss' least effort
principle.  This means equations of motion like
\begin{eqnarray}
&&m \,\ddot{\V x}_j^0= -\partial_{\V x_j^0} \big(V(\V X^0) +\sum_{a=1}^{M}
W_a(\V X^0,\V X^a)\big)+\V F_i(\V X^0,\BF)\nonumber\\
&&m_a\, \ddot{\V x}^a_j=  -\partial_{\V x^a_j} \big( V_a(\V X^a)+W_a(\V X^0,\V
X^a)\big)-\Bth^a_j
\label{3.1}\end{eqnarray}
and an application of Gauss' principle yields $\Bth_j^a=
\frac{m_a(L_a-\dot V_a)} {(dN_a-1)
k_B T_a}\,\,\dot{\V x}^a_j\defi \a^a \,\dot{\V
x}^a_j $ where $L_a$ is the work per unit time done by the particles
in $\Si_0$ on the particles of $\Si_a$ and $V_a$ is their potential
energy, \cite{Ga05}. Note that in the first of Eq. (\ref{3.1}), the forces 
$-\partial_{\V x_i^0}W_a$ play the role of the thermostat forces $\Bth_i$ in 
Eq. (\ref{1.1}). 

The work $L_a$ appearing in the definition of $\Bth^a_j$
can be naturally interpreted as {\it heat} $\dot
Q_a$ ceded, per unit time, by the particles in $\Si_0$ to the
thermostat $\Si_a$ (because the ``temperature'' of $\Si_a$ is constant).
If $\V X=(\V X^0,\V X^1,\ldots,\V X^M)$, 
the {\it entropy creation rate} can be naturally defined as
\begin{eqnarray}\s_0(\dot{\V X},\V X)\defi\sum_{a=1}^{M}
\frac{\dot Q_a}{k_B T_a}\label{3.2}\end{eqnarray}
hence $\s_0$ can be called (in model Eq. (\ref{3.1}))
the {\it average entropy creation rate}. Its time average
will be assumed $\s_{0,+}\defi\media{\s_0}_{SRB}\ne0$. Note that
now $\media{\cdot}_{SRB}$ is the average with respect to the 
stationary measure for the whole system $\Si_0$+thermostats.

We shall see that, in ample generality, $\s_{0+}\ge0$: the definition
of entropy creation is ``reduced'', here, to an ``equilibrium notion''
because what is being defined is the entropy increase of the
thermostats, which have to be considered in equilibrium. No attempt is
made to define the entropy of the stationary state. Nor any attempt is
made to define the temperature of the nonequilibrium system in $\Si_0$
($T_a$ is the temperature of $\Si_a$, not of $\Si_0$).  

In fact the above model is a realization of a {\it Carnot
machine}: the machine being the system in $\Si_0$ on which external
forces work leaving the system in the same state (a special ``cycle'')
but achieving a transfer of heat between the various thermostats (in
agreement with the second law, see Eq. (\ref{3.2}), only if
$\s_{0,+}\ge0$).

Another observable that is convenient to introduce is the {\it phase space
contraction rate} $\s(\dot {\V X},\V X)$, defined as 
{\it minus the divergence} of the equations of motion:
in the model described by Eq. (\ref{3.1}), this
is the sum of the derivatives of the {\it r.h.s.} with respect to 
$\V X^0, \V X^a, m\dot{\V X}^0, m_a\dot{\V X}^a$ and it turns out to be
\begin{eqnarray}\s(\dot{\V X},\V X)=\sum_a
\frac{\dot Q_a-\dot V_a}{k_B T_a}\=
\sum_a \frac{\dot Q_a}{k_B T_a}-\dot u\;.\label{5.1}\end{eqnarray}
Therefore there is a simple and direct relation between the phase
space contraction and the entropy creation rate, \cite{Ga05}: they just differ 
for the total derivative $\dot u\defi\sum_a \frac{ \dot V_a}{k_B T_a}$,
whose time average $\media{\dot u}_{SRB}$ vanishes. This implies
in particular that the time averages $\s_+\defi\media{\s}_{SRB}$
is the same as $\s_{0,+}$ and in particular is nonnegative, consistently
with the interpretation of phase space volume contraction
rate. The $\s_+>0$ implies that
phase space contracts in average and therefore the SRB distribution
will give probability $1$ to a zero volume set. Therefore if $\s_+>0$
the system is said {\it dissipative}.

The usefulness of introducing the definition of $\s$ is that a {\it
fluctuation theorem} has been established for its large deviations functional
in the context of Anosov systems theory.
The {\it chaotic hypothesis} allows to establish a connection between the 
fluctuation theorem (valid for the phase space contraction rate in 
dissipative Anosov systems) and a {\it fluctuation relation} 
for $\s$ and $\s_0$ in model Eq. (\ref{3.1}), as explained in next sections.

\section{Chaotic hypothesis}

In equilibrium statistical mechanics the {\it ergodic hypothesis}
plays an important conceptual role as it implies that the motions have
a SRB statistics and that the latter coincides with the Liouville
distribution on the energy surface.  A role analogous to the {\it
ergodic hypothesis} has been proposed for the {\it chaotic
hypothesis}, \cite{GC95}: which states that

\*
\noindent{}{\it A chaotic system can be regarded as an Anosov system
to the purpose of computing the time averages of (smooth) observables.}
\*

This means that the attracting set of a chaotic system, physically
defined as a system with at least one positive Lyapunov exponent, can
be regarded as a smooth compact surface $\O$ on which motion is highly unstable
(uniformly hyperbolic) and transitive (there is a dense trajectory).
For a mathematically precise definition of Anosov system we refer to 
\cite{GBG04}.

\kern2pt
We stress that the chaotic hypothesis concerns physical
systems: mathematically {\it it is easy to find dynamical systems
for which it does not hold}. As it is easy (even easier) to
find systems in which the ergodic hypothesis fails ({\it e.g.}
harmonic lattices or black body radiation).

Since physical systems are almost always not Anosov systems it is very
likely that probing motions in extreme regimes will make visible the
features that distinguish Anosov systems from non Anosov systems: much
as it happens with the ergodic hypothesis. 

The ergodic hypothesis provides us with an expression for the averages
(as integrals over the normalized Liouville distribution on the energy
surface): likewise the Chaotic Hypothesis provides us with the
existence and a formal expression for the averages ({\it i.e.} for the SRB
distribution), \cite{GBG04}.

The interest of the hypothesis is to provide a framework in which
properties like existence and formal expression of an SRB
distribution is {\it a priori} guaranteed. one can also say that the role of
Anosov systems in chaotic dynamics is similar to the role of harmonic
oscillators in the theory of regular motions.  They are the paradigm
of chaotic systems as the harmonic oscillators are the paradigm of
order. Of course the hypothesis is only a beginning and one has to
learn how to extract information from it, as it was the case with the
use of the Liouville distribution once the ergodic hypothesis
guaranteed that it was the appropriate distribution for the study of
the statistics of motions in equilibrium situations, \cite{BGGZ05}.

\section{Fluctuation theorem}

As mentioned above, an important observable in the theory of Anosov systems is 
the phase space contraction rate $\s(x)$, defined as minus the divergence
of the equations of motion, computed in $x\in\O$, where $\O$ is the phase space
of the system. 
A rather general result holds {\it if} the system is Anosov,
dissipative ($\s_+=\media{\s}_{SRB}>0$), 
and {\it furthermore} reversible in the sense
that there is an isometry $I$ of phase space such that $I S_t=S_{-t}I$
for all $t\in\RRR$.  Define the {\it dimensionless phase space
contraction}
\begin{eqnarray}p(x)=\frac1\t 
\int_0^\t \frac{\s(S_t x) }{\s_+}\, dt\label{4.2}\end{eqnarray}
then the SRB average of $p$ is $1$ (by definition) and there exists
$p^*\ge1$ such that the probability $P_\t$ of the event $p\in [a,b]$
with $[a,b]\subset (-p^*,p^*)$ has the form
\begin{eqnarray}P_\t(p\in[a,b])\,=\,const\, e^{\t \max_{p\in [a,b]} \z(p)
+O(1)}\label{4.3}\end{eqnarray}
with $\z(p)$ analytic in $(-p^*,p^*)$, \cite{Si72,Si94}.  The function
$\z(p)$ can be conveniently normalized to have value $0$ at $p=1$ ({\it i.e.}
at the average value of $p$).

{\it In Anosov systems which are reversible and dissipative} a
general symmetry property, called the {\it fluctuation theorem} and
reflecting the reversibility symmetry, yields the {\it parameterless}
relation, \cite{GC95,Ge98},
\begin{eqnarray}\z(-p)=\z(p)-p\s_+ \qquad p\in(-p^*,p^*)\;,
\label{4.4}\end{eqnarray}
where $(-p^*,p^*)$, $p^*\ge 1$, is the largest domain of definition of $\z$;
it can be shown that $\z$ is analytic on the whole $(-p^*,p^*)$.
This relation is interesting because it has no free parameters. The
relation was discovered in a simulation of a shear flow and it was
suggested that it should be related to time reversal symmetry and to
Ruelle's ideas on turbulence, \cite{ECM93}.  A more informal (but
imprecise) way of writing Eq. (\ref{4.3}),(\ref{4.4}) is
\begin{eqnarray}\frac{P_\t(p)}{P_\t(-p)}=e^{\t p \s_++ O(1)}, 
\qquad \ {\rm for\ all} \
p\in(-p^*,p^*)
\label{4.5}\end{eqnarray}
where $P_\t(p)$ is the probability density of $p$. An interesting
consequence of Eq. (\ref{4.5}) is that $\media{e^{-\t\,p\,\s_+}}_{SRB}=1$ in
the sense that $\frac1\t \log
\media{e^{-\t\,p\,\s_+}}_{SRB}\tende{\t\to\infty}0$.

Occasionally systems with singularities have to be considered: in such
cases the relation Eq. (\ref{4.4}) may change in the sense that the 
$\z(p)$ may be not analytic; one then expects that the relation holds in
the largest analyticity interval symmetric around the origin. In
various cases considered in the literature such interval appears to
contain the interval $(-1,1)$ and sometimes this can be proved
rigorously. For instance in simple, although admittedly special,
examples of systems close to equilibrium, \cite{BGGZ05}.

The Eq. (\ref{4.4}),(\ref{4.5}) is the first representative of
consequences of the reversibility and chaoticity
hypotheses. For instance given $F_1,\ldots,F_n$ {\it arbitrary}
observables which are (say) odd under time reversal $I$ ({\it i.e.} $F(I
x)=-F(x)$) and given $n$ functions $t\in[-\frac\t2,\frac\t2]\to
\f_j(t)$, $j=1,\ldots,n$ one can ask which is the probability that
$F_j(S_tx)$ ``closely follows'' the {\it pattern} $\f_j(t)$ and at the
same time $\frac1\t\int_0^\t \frac{\s(S_\th x)}{\s_+}\,d\th$ has value
$p$. Then calling $P_\t(F_1\sim\f_1,\ldots,F_n\sim\f_n,p)$ the
probability of this event, which we write in the imprecise form
corresponding to Eq. (\ref{4.5}) for simplicity, and defining
$I\f_j(t)\defi-\f_j(-t)$ it is
\begin{eqnarray}\frac{P_\t(F_1\sim\f_1,\ldots,F_n\sim\f_n,p)}
{P_\t(F_1\sim I\f_1,\ldots,F_n\sim I\f_n,-p)}
=e^{\t \s_+ p},\label{4.6}\end{eqnarray}
for $p\in (-p^*,p^*)$, which is remarkable because it is parameterless
and at the same time surprisingly independent of the choice of the
observables $F_j$.  The relation Eq. (\ref{4.6}) has far reaching
consequences, \cite{Ga00}.

Eq. (\ref{4.6}) can be read as follows: the probability that the
observables $F_j$ follow given {evolution patterns} $\f_j$ conditioned
to entropy creation rate $p\s_+$ is {\it the same} that they follow
the time reversed patterns if conditioned to entropy creation rate
$-p\s_+$. In other words to change the sign of time it is {\it just}
sufficient to reverse the sign of the average phase space
contraction $p$ (and we shall see that in our model this would amount
at changing the sign of the entropy creation rate): no ``extra effort''
is needed.

\section{Fluctuation relation}

Given a chaotic system for which the Chaotic Hypothesis can be
regarded as valid it is expected that the Fluctuation Relation (FR) holds: 
{\it i.e.} that one
can define $\z(p)$ and that the symmetry Eq. (\ref{4.4}) holds.
This is an important check that can be performed on the statistical
properties of a stationary nonequilibrium state.

It is however also important to know whether the quantity $p$ and its
fluctuations describe some interesting feature of the dynamical
system. The model Eq. (\ref{3.1}), with $\s$ defined as in Eq. (\ref{5.1}), 
provides an indication on the path to
follow in the quest of an interpretation of the Fluctuation Relation \cite{Ga06,Ga05}.

The remarkable property is that if we accept the Chaotic Hypothesis 
for the model Eq. (\ref{3.1}) ({\it i.e.} for the whole system 
$\Si_0$+thermostats) and we choose the parameters $\BF$ and $T_a$ 
in such a way that $\s_+>0$, it is
expected that FR holds for $p$: and this has an
immediate physical interpretation because
if we write $p_0(x)\defi\frac1\t \int_0^\t \frac{\s_0(S_t x) }{\s_{0+}}\,
dt$ then, making use of the property that $p-p_0$ is the variation of
$u$ in time $\t$ divided by the elapsed time $\t$:
\begin{eqnarray}\label{5.2}
p(x)=p_0(x)+ \frac{u(S_tx)-u(x)}t\cr
\s_{0+}\=\media{\s_0}_{SRB}\=\media{\s}_{SRB}\=\s_+
\end{eqnarray}
we find that in the limit $t\to\infty$ $p$ and $p_0$ have {\it the same large
deviation rate $\z(p)$}.

Of course the thermostats are ``large'' (we are even neglecting
$O(N_a^{-1})$) and therefore the energies $V_a$ as well as $u$ can be
very large, physically of order $O(\sum_a N_a)$. This means that the
time $t$ that has to pass to see the fluctuation relation Eq. (\ref{4.4})
free of the $O(t^{-1}\sum_a N_a)$ corrections can be enormous
(possibly on astronomical scale in ``realistic'' cases).

This is a situation similar to the one met when considering systems
with unbounded stochastic forces, \cite{CV03a}, or with singular or
nearly singular forces and thermostats of isokinetic type,
\cite{BGGZ05}. In the present case we see that the really interesting
quantity is the quantity $p_0$: which is the entropy creation rate and
it is a boundary term {\it unaffected by the size of the
thermostats}. Therefore if one considers and measures only $p_0$ rather
than $p$ one not only performs a physically meaningful operation ({\it i.e.}
measuring the average entropy creation rate) but also one can access
the large deviation rate $\z(p)$ and check the Fluctuation Relation
symmetry on a time totally unrelated to the thermostats size.
Furthermore the more general relations like Eq. (\ref{4.6}) can be naturally
extended.

\section{A model for granular materials}

The current interest in granular materials properties and the
consequent availability of experiments, {\it e.g.} \cite{FM04}, suggests
trying to apply the above ideas to derive possible experimental tests.

The main problem is that in granular materials collisions
are intrinsically {\it inelastic}. In each collision particles heat up,
and the heat is subsequently released trough thermal exhange with the walls of
the container, sound emission (if the experiment is performed in air),
radiation, and so on. If one still wants to deal with a {\it reversible}
system, such as the one we discussed in the previous sections, one should
include all these sources of dissipation in the theoretical description.
Clearly, this is a {\it very} hard task, and we will not pursue it here.

A simplified description of the system consists in neglecting the internal
degrees of freedom of the particles. In this case the inelastic collisions
between point particles will represent the only source of dissipation
in the system. Still the chaotic hypothesis is expected to hold,
but in this case the entropy production is strictly
positive and there is no hope of observing a fluctuation relation,
see {\it e.g.} \cite{PVBTW05}, if one looks at the whole system.

Nevertheless, in the presence of inelasticity, temperature gradients
are present in the system \cite{GZN96,BMM00,FM04}, so that heat is
transported through different regions of the container.
Then one can try to represent the processes of heat exchange between different
regions of the system by the model we described above: and assuming that,
under suitable conditions, the inelasticity of the collisions can be
neglected, one can hope to observe a fluctuation relation for a (suitably
defined) entropy production rate. 
This would be an interesting example of ``ensemble equivalence'' in
nonequilibrium \cite{Ga00}: we will discuss this possibility in
detail in the following.


As a model for a granular material let $\Si$ be a container consisting
of two flat parallel vertical walls covered at the top and with a
piston at the bottom that is kept oscillating by a motor so that its
height is
\begin{eqnarray}z(t)= A \cos\o t\label{6.1}\end{eqnarray}
The model can be simplified introducing a sawtooth 
moving piston as in \cite{BMM00}, however the results should not 
depend too much on the details of the time dependence of $z(t)$.
The container $\Si$ is partially filled with millimiter size balls
(a typical size of the faces of $\Si$ is $10\ cm$ and the particle number
is about $256$): the vertical walls are so close that the balls almost
touch both faces so the problem is effectively two dimensional.
The equations of motion of the balls with
coordinates $(x_i,z_i), \, i=1,\ldots,N$, $z_i\ge z(t)$, are
\begin{eqnarray}
m \ddot{x}_i= f_{x,i}&\hfill\label{6.2}\cr 
m \ddot{z}_i= f_{z,i}& -m g + m\d(z_i-z(t)) \, 2\, (\dot z(t)-\dot
z_i)\end{eqnarray}
where $m$=mass, $g$=gravity acceleration, and the collisions
between the balls and the oscillating base of the container are assumed
to be elastic \cite{BMM00} (eventually inelasticity of the walls can be
included into the model with negligible changes \cite{PVBTW05}); 
$\V f_i$ is the force describing the particle collisions and the 
particle-walls collisions.

The force $\V f_i$ has a part describing the particle collisions:
this is not necessarily elastic and in fact we will assume 
that the particle collisions are inelastic, with restitution coefficient 
$\a<1$. A simple model for inelastic collisions with inelasticity $\a$
(convenient for numerical implementation)
is a model in which collisions take place with the usual elastic collision 
rule but immediately after the velocities of the particles that have collided
is scaled by a factor so that the kinetic energy of the pair is
reduced by a factor $1-\a^2$ \cite{GZN96,BMM00,PVBTW05}.

We look at the stationary distribution of the balls: the simplest
experimental situation that seems accessible to experiments and
simulations is to draw ideal horizontal lines at heights $h_1>h_2$
delimiting a strip $\Si_{0}$ in the container and to look at the
particles in $\Si_{0}$ as a thermostatted system, the thermostats
being the regions $\Si_1$ and $\Si_2$ at heights larger than $h_1$
and smaller then $h_2$, respectively. 

After a stationary state has been reached, the average kinetic energy
of the particles will depend on the height $z$, and in particular will
decrease on increasing $z$.
Given the motion of the particles and a time interval $t$ it will 
be possible to measure the quantity $Q_2$ of (kinetic) energy that 
particles entering or exiting the region $\Si_{0}$ from below 
(the ``hotter side'') carry out of $\Si_{0}$ and the analogous quantity 
$Q_1$ carried out by the particles that enter or exit from above 
(the ``colder side'').

If $T_i,\,i=1,2$ are the average
kinetic energies of the particles in small horizontal corridors
above and below $\Si_{0}$, we see that there 
is a connection between the model of granular material, Eq.~(\ref{6.2}), and 
the model Eq.~(\ref{3.1}) discussed above. Still, model Eq.~(\ref{6.2})
cannot be exactly reduced to model Eq.~(\ref{3.1}) because of the 
internal dissipation induced by the inelasticity $\a$ and of the fact that 
the number of particles in $\Si_0$ depends on time, as particles come and go in
the region. 

The reason for considering a model of granular material that is not
in the class of models Eq. (\ref{3.1}) is that Eq. (\ref{6.2}) has 
a closer connection with the actual experiments \cite{FM04} and with the 
related numerical simulations. Moreover, under suitable assumptions, 
that can be expected to hold on a specific time scale, 
the stationary state of Eq. (\ref{6.2}) is effectively described 
in terms of the stationary state of Eq. (\ref{3.1}), as discussed below.

Note that real experiments cannot have an arbitrary duration, \cite{FM04}: the
particles movements are recorded by a digital camera and the number of
photograms per second is of the order of a thousand so that the memory
for the data is easily exhausted as each photogram has a size of about
a $1$Mb in current experiments. The same holds for numerical simulations
where the accessible time scale is limited by the available computational
resources.

Hence each experiment lasts up to a few seconds starting after the
system has been moving for a while so that a stationary state is
reached. The result of the experiment is the reconstruction of the
trajectory in phase space of each individual particle inside the
observation frame, \cite{FM04}.

In order for the number of particles $N_0$ in $\Si_0$ to be approximately 
constant for the duration of the experiment, the vertical size $(h_1-h_2)$ of 
$\Si_0$ should be chosen large compared to $(Dt)^{1/2}$, where $t$ 
is the duration of the experiment and $D$ is the diffusion coefficient. Note 
that we are assuming that the motion of the particles 
is diffusive on the scale of $\Si_0$. In the low density case the motion could 
be not diffusive on the scale of $\Si_0$: then we would not be able 
to divide the degrees of freedom between the subsystem and the rest of the 
system and moreover the correlation length would be comparable with 
(or larger than) the size of the subsystem $\Si_0$. This would completely 
change the nature of the problem and violations to FR could possibly 
be observed \cite{BCL98,BL01}.

Given the remarks above and if
\*

\noindent{}(1) we accept the chaotic hypothesis,

\noindent{}(2) we assume that the result of the observations would be 
the same if the particles above $\Si_{0}$ and below $\Si_{0}$ were kept at
constant total kinetic energy by reversible thermostats ({\it e.g.} 
Gaussian thermostats), \cite{ES93,Ga00,Ru00},

\noindent{}(3) we neglect the dissipation due to inelastic collisions between
particles in $\Si_{0}$,

\noindent{}(4) we neglect the fluctuations of the number of particles in 
$\Si_0$,

\noindent{}(5) we suppose that there is dissipation in the sense that
\begin{eqnarray}\s_+\defi \lim_{t\to+\infty}
\frac1t\Big(\frac{Q_1}{T_1}+\frac{Q_2}{T_2}\Big)>0\;,\label{6.3}\end{eqnarray}
we expect the analysis of Sect.5 to apply to model Eq.~(\ref{6.2}). 

Note that chaoticity is expected at least if dissipation is small and evidence
for it is provided by the experiment in \cite{FM04} which indicates
that the system evolves to a chaotic stationary state in which
dissipation occurs. Dissipation due to internal inelastic collisions
will be negligible (to the purpose of checking a FR for $\s_0$)
only on a specific time scale, as discussed below.

Accepting the assumptions above, we then predict that a Fluctuation Relation
is satisfied, see Eq. (\ref{4.4}) and (\ref{4.5}), for fluctuations of
\begin{eqnarray}p=\frac1{t\,\s_+}{\Big(\frac{Q_1}{T_1}+\frac{Q_2}{T_2}\Big)}
\label{6.4}\end{eqnarray}
in the interval $(-p^*,p^*)$ with $p^*$ equal (at least) to $1$.

The latter is therefore a property that seems accessible to simulations as 
well as to experimental test. Note however that it is very likely that the 
hypotheses (2)-(4) above will not be {\it strictly} verified in real
experiments, see the discussion in next section, so we expect that the 
analysis and interpretation of the experimental results will be non trivial. 
Nevertheless, the test would be rather stringent.

\section{Relevant time scales}

The above analysis assumes the existence of (at least) two time scales. 
One is the ``equilibrium time scale'', $\th_e$, which is the time scale
over which the system evolving at constant energy, equal to the average
energy observed, would reach equilibrium in absence of friction and
forcing. An experimental measure of $\th_e$ would be the decorrelation
time of self--correlations in the stationary state, and 
we can assume that $\th_e$ is of the order of the mean collision time.
Note that $\th_e$ also coincides with the time scale over which 
finite time corrections to FR become irrelevant \cite{ZRA04}: this means that 
in order to be able to measure the large deviations functional for the 
normalized entropy production rate 
$p$ in Eq.~(\ref{6.4}) one has to choose $t\gg\th_e$, see 
also \cite{GZG05} for a detailed discussion of the first orders finite time
corrections to the large deviation functional.
A second time scale is the ``inelasticity time scale'' $\th_d$, which is 
the scale over which the system reaches a stationary state if the particles 
are prepared in a 
random configuration and the piston is switched on at time $t=0$.
Possibly a third time scale is present: the ``diffusion time 
scale'' $\th_{D}$ which is the scale over which a particle diffuses
over the size of $\Si_0$.
The analysis above applies only if the time $t$ in Eq.~(\ref{6.4})
verifies $\th_e\ll t \ll\th_d, \th_D$ (note however that the measurement 
should be started
after a time $\gg \th_d$ since the piston has been switched on in order to
have a stationary state); in practice this
means that the time for reaching the stationary state
has to be quite long compared to $\th_e$. In this
case friction is negligible for the duration of the measurement if the
latter is between $\th_e$ and $\min(\th_D,\th_d)$.
In the setting we consider, the role of 
friction is ``just'' that of producing the nonequilibrium stationary state 
itself and the corresponding gradient of temperature: this is reminiscent of 
the role played by friction in classical mechanics problems, where periodic 
orbits (the ``stationary state'') can be dynamically selected by adding a 
small friction term to the Hamilton equations. Note that, as discussed below, 
the temperature gradient produced by friction will be rather small: however
smallness of the gradient does not affect the ``FR time scale'' over which
FR is observable \cite{ZRA04}. 

If internal friction is not negligible (that is if $t\gtrsim \th_d$)
the problem would change nature: an explicit model (and theory) should be 
developed to describe the transport mechanisms (such as radiation, 
heat exchange between the particles and the container, sound emission, ...) 
associated with the dissipation of kinetic energy and new thermostats should 
be correspondingly introduced. The definition of entropy production 
should be changed, by taking into account the presence of such new 
thermostats. In this case, even changing the definition of entropy production 
it is not clear whether FR should be satisfied: in fact internal dissipation
would break the time--reversibility assumption and, even accepting the Chaotic 
Hypothesis, nothing guarantees a priori the validity of FR.
 
The validity of $\th_e\ll t \ll\th_d, \th_D$ is not obvious in experiments.
A rough estimate of $\th_d$ can be given as follows: the phase space
contraction in a single collision is given by $1-\a$. Thus the average
phase space contraction per particle and per unit time is
$\s_{+,d} = (1-\a)/\th_e$, where $1/\th_e$ is the frequency of the collisions
for a given particle. It seems natural to assume that $\th_d$ is the time
scale at which $\s_{+,d} \th_d$ becomes of order $1$: on this time scale
inelasticity will become manifest. Thus, we obtain the following estimate:
\begin{equation}\label{thd}
\th_d \sim \frac1{1-\a} \th_e
\end{equation}
In real materials $\a\le .95$, so that $\th_d$ can be at most
of the order of $20 \th_e$. Nevertheless it is possible that this is 
already enough to observe a Fluctuation Relation on intermediate times.

The situation is completely different in numerical simulations where
we can play with our freedom in choosing the restitution coefficient $\a$ 
(it can be chosen very close to one \cite{GZN96,BMM00,PVBTW05},
in order to have $\th_d\gg\th_e$) 
and the size of the container $\Si_0$ (it can be chosen large,
in order to have $\th_D\gg\th_e$).

To check the consistency of our hypotheses, it has to be
shown that it is possible to make a choice of parameters so that 
$\th_d$ and $\th_D$ are separated by a large time window.
Such choices are possible, as discussed below.

If $\d=h_1-h_2$ is the width of $\Si_0$,
$\e=1-\a$, $\g$ is the temperature gradient in $\Si_0$, and $D$ the
diffusion coefficient the following estimates hold:
\*

\noindent(a) $\th_e=O(1)$ as it can be taken of the order of the inverse
collision frequency, which is $O(1)$ if density is constant and 
the forcing on the system is tuned to keep the energy constant as $\e\to0$. 
\\
(b) $\th_d=\th_e O(\e^{-1})$ as implied by Eq.~(\ref{thd}).
\\
(c) $\th_D=O(\frac{\d^2}D)=O(\d^2)$ because $D$ is a constant
(if the temperature and the density are kept constant).
\\
(d) $\g=O(\sqrt\e)$, as long as $\d\ll\e^{-1/2}$. 
In fact if the density is high enough
to allow us to consider the granular material as a fluid, as in
Eq.~(5) of Ref.\cite{BMM00}, the temperature profile should be given
by the heat equation $\nabla^2 T+c\e T=0$ with suitable constant $c$
and suitable boundary conditions on the piston ($T=T_0$) and on the top of 
the container ($\nabla T=0$). 
This equation is solved by a linear combination of
$const \,e^{\pm\sqrt{c\e} z}$, which has
gradients of order $O(\sqrt\e)$, as long as $\d \ll 1/\sqrt\e$ and the
boundaries of $\Si_0$ are further than $O(1/\sqrt\e)$ from the top.

Now, if we choose $\d=\e^{-\b}$, with $\b<\frac12$,
and we take $\e$ small enough, we have $\th_e \ll \min\{\th_d,\th_D\}$ and
$\d\ll O(\e^{-\frac12})$, as required by item (d). 
\\

\noindent{\it Remark:} 
The entropy creation rate due to heat transport into $\Si_0$, in
presence of a temperature gradient $\g$, is given by $\s_+=O(\g^2
\d)=O(\e\d)$ because the temperature difference is $O(\g\d)$ and the energy
flow through the surface is of order $O(\g)$ (with $\g=O(\sqrt\e)$, see 
item (d)). The order of magnitude of $\s_+$ is not larger then the
average amount $\s_d$ of energy dissipated per unit time in $\Si_0$ divided 
by the average kinetic energy $T$ (the latter quantity is of order
$O(\th_e^{-1}\e\d)$ because, at constant density, the number of
particles in $\Si_0$ is $O(\d)$); however the entropy creation due to
the dissipative collisions in $\Si_0$ has fluctuations of order
$O(\e\d^{\frac12})$ because the number of particles in $\Si_0$
fluctuates by $O(\d^{\frac12})$. This is consistent with neglecting the entropy
creation inside the region $\Si_0$ due to the inelasticity in spite of
it being of the same order of the entropy creation due to the heat
entering $\Si_0$ from its upper and lower regions.\\

\*

This argument supports the proposal that in numerical simulations
it will be possible to test our ideas by a suitable choice of the 
parameters. We expect that other choices will be possible: for instance
in the high-density limit it is clear that $\th_D \gg \th_e$ because
the diffusion coefficient will become very small.
To what extent this can be applied to experiment remains 
an open question.

\section{Remarks and conclusions}

\noindent
(1) The model can be given further structure by
adding a non conservative forcing acting on the particles in the
region $\Si_{0}$: the same relations would follow (in particular
the Fluctuation Relation) if the forced equation of motion are still
reversible; see \cite{ZBCK05} for a (stochastic) example.
The above will not hold in general if the forcing is not reversible, 
{\it e.g.} if the inelasticity of the collisions inside $\Si_{0}$ cannot be
neglected, see below.
\\ 
(2) An explicit computation of the large deviation function of the
dissipated power, in the regime $t \gg \th_d$ ({\it i.e.} when the dissipation
is mainly due to inelastic collisions) recently appeared in 
\cite{VPBTW05}. However in the model only the dissipation due to the
collisions was taken into account: so that it is not clear how the heat
produced in the collisions is removed from the system, see the discussion
above. It turned out that in this regime no negative
values of $p$ are observed so that the FR cannot hold. This is interesting
and expected on the basis of the considerations above. It is not clear
if, including the additional thermostats required to remove heat from the
particles and prevent them to warm up indefinitely, the Fluctuation 
Relation is recovered. However this problem is different from the one
discussed here and we leave it for future investigation.
\\
(3) There has also been some debate on the interpretation of the experimental
results of \cite{FM04}. In \cite{PVBTW05} a simplified model, very similar
to the one discussed above, was proposed
and showed to reproduce the experimental data of \cite{FM04}. The prediction
of the model is that the FR is not satisfied. Note however that the geometry 
considered in
\cite{FM04,PVBTW05} is different from the one considered here: the whole
box is vibrated, so that the the temperature profile is symmetric, and
a region $\Si_0$ 
in the center of the box is considered. Heat exchange is due to
``hot'' particles extering $\Si_0$ ($Q_+$) and ``cold''
particles exiting $\Si_0$ ($Q_-$). One has
$Q=Q_+ + Q_- \neq 0$ because of the dissipation in $\Si_0$ and
\begin{eqnarray}\s_+ = \frac{\dot Q_+}{T_+} + 
\frac{\dot Q_- }{T_-} =0 \label{6.5}\end{eqnarray}
where $T_+$ is the temperature outside $\Si_0$ and $T_-$ is the
temperature inside $\Si_0$, see \cite{PVBTW05}.
Thus the dissipation due to heat exchange between the region
$\Si_{0}$ and the regions outside $\Si_{0}$ vanishes and the only
dissipation is due to the inelasticity of the collisions. In this
regime, again, the FR is not expected to hold if the thermostat
dissipating the heat produced in the collisions is not included in the
model, see above: it is an interesting remark of \cite{PVBTW05} that
partially motivated the present work.
We believe that different experiments
can be designed in which the dissipation is mainly due to heat
exchanges and the inelasticity is negligible, as the one we proposed
above.  The main difference between the experiment we proposed and the
one in \cite{FM04} is that the geometry of the box is such that
$Q_1/T_1+Q_2/T_2 > 0$: in this situation the
dissipation due to inelastic collisions should be negligible as long
as $t \ll \th_d$.
\\
(4) Even in situations in which the dissipation is entirely 
due to irreversible
inelastic collisions between particles, such as the ones considered in
\cite{PVBTW05,VPBTW05}, the Chaotic Hypothesis is expected to hold,
and the stationary state to be described by a SRB distribution.
In these cases the failure of the Fluctuation Relation 
is not in contradiction with the chaotic hypothesis, 
due to the irreversibility of the equations of motion.  

\*
\noindent{\it Conclusions} We showed that in a paradigmatic class of mechanical models of
thermostatted systems the phase space contraction is an interesting
quantity. In large systems in contact with thermostats it may consist of a sum 
of two quantities, the first with the interpretation of entropy creation rate 
and the second extremely large but equal to a total derivative.

Its fluctuation properties, while asymptotically for large
times determining (actually being identical to) the fluctuation
properties of the entropy creation rate, {\it hence implying the
fluctuation relation}, may require very long time to be freed of
finite time corrections. But at the same time the study of the
fluctuation properties of the physical quantity defined by the entropy
creation rate can be used to determine the large deviations of the
phase space contraction. The latter, having the {\it same} large
deviation rate, must obey the Fluctuation Relation which therefore
becomes observable even if the system is in contact with large (or
even infinite) thermostats.

The analysis leads to propose concrete experimental tests as well as
tests based on simulations in the context of granular materials.
The models naturally introduced for the description of granular materials
experiments are not in the same class of models for which a 
relation between phase space contraction and entropy production rate was
previously discussed. The previous analysis can be applied to granular 
materials only under suitable assumptions, verified on a specific 
time scale. A fluctuation relation for an entropy production rate measured 
on the same time scale is predicted.

\acknowledgments

We have profited of several discussions and
comments from A.~Barrat, S.~Fauve, T.~Gilbert, A.~Puglisi, P.~Visco and
F.~van~Wijland.
A.G. was partially supported by U.S. National Science Foundation
grant PHY 01 39984.
F.Z.~has been supported by the EU Research Training Network STIPCO
(HPRN-CT-2002-00319).

\bibliographystyle{apsrev} 

\revtex
\end{document}